%% file: GW_PMF_revision.tex
\documentclass[aps, prd, twocolumn, nofootinbib, floatfix, superscriptaddress]{revtex4-1}

\usepackage{graphicx}% Include figure files
\usepackage{dcolumn}% Align table columns on decimal point
\usepackage{bm}% bold math
\usepackage{subfigure}
\usepackage{epsfig}
\usepackage{amsmath}
\usepackage{amsfonts}
\usepackage{graphicx}
\usepackage{amssymb}
\usepackage{amssymb,amsmath,amsfonts}
\usepackage{xfrac}
\usepackage{color}
\usepackage{tikz}
\usepackage[T1]{fontenc}
\usepackage[utf8]{inputenc}

\usepackage{tcolorbox}
\usepackage{hyperref}
\hypersetup{colorlinks, citecolor=bluscuro, linkcolor=black, urlcolor=bluscuro}
\definecolor{rossos}{cmyk}{0,1,1,0.55}
\definecolor{bluscuro}{rgb}{0.15, 0.2, .85}
\definecolor{bluchiaro}{cmyk}{1,.3,0.,0.1}

\input{newcommands}

\newcommand{\be}{\begin{equation}}

\def\apj{ApJ}
\def\mnras{MNRAS}

\def\prd{Phys. Rev. D}
\def\aap{A\&A}                   % "Astron. Astrophys."
\def\apjl{ApJ}                   % letter at ApJ

\begin{document}

\title{Constraining supermassive primordial black holes with magnetically induced gravitational waves}% Force line breaks with \\

\author{Theodoros Papanikolaou}
\email{papaniko@noa.gr}
 \affiliation{National Observatory of Athens, Lofos Nymfon, 11852 Athens, 
Greece}%Lines break automatically or can be forced with \\
\author{Konstantinos N. Gourgouliatos}%
 \email{kngourg@upatras.gr}
\affiliation{Department of Physics, University of Patras, Patras, Rio, 26504, Greece}%

%\date{}% It is always \today, today,
             %  but any date may be explicitly specified

\begin{abstract} 
Primordial black holes (PBHs) can answer a plethora of cosmic conundra, among which the origin of the cosmic magnetic fields. In particular, supermassive PBHs with masses $M_\mathrm{PBH}>10^{10} M_\odot$ and furnished with a plasma-disk moving around them can generate through the Biermann battery mechanism a seed primordial magnetic field which can later be amplified so as to provide the magnetic field threading the intergalactic medium. In this work, we derive the gravitational wave (GW) signal induced by the magnetic anisotropic stress of such a population of magnetised PBHs. Interestingly enough, by using GW constraints from Big Bang Nucleosynthesis (BBN) and an effective model for the galactic/turbulent dynamo amplification of the magnetic field, we set a conservative upper bound constraint on the abundances of supermassive PBHs at formation time, $\Omega_\mathrm{PBH,f}$ as a function of the their masses, namely that $\Omega_\mathrm{PBH,f}\leq 2.5\times 10^{-10}\left(\frac{M}{10^{10}M_\odot}\right)^{45/22}$. Remarkably, these constraints are comparable, and, in some mass ranges, even tighter compared to the constraints on $\Omega_\mathrm{PBH,f}$ from large-scale structure (LSS) probes; hence promoting the portal of magnetically induced GWs as a new probe to explore the enigmatic nature of supermassive PBHs.
\end{abstract}

%Use showkeys class option if keyword
                              %display desired
\maketitle

%\tableofcontents
%%%%%%%%%%%%%%%%%%%%%%%%%%% SECTION 1: Introduction 
%%%%%%%%%%%%%%%%%%%%%%%%%%%%%%%%%%%

\section{Introduction} 
The origin of the primordial magnetic fields (MFs) threading the intergalactic medium constitutes one of the longstanding issues in cosmology. These cosmic MFs can play a crucial role in the processes of particle acceleration through the intergalactic medium~\cite{Bagchi:2002vf}, of the propagation of cosmic rays~\cite{Strong:2007nh} while at the same time, they can affect significantly the Universe's thermal state between inflation and recombination~\cite{Grasso:1996kk,Barrow:1997mj,Subramanian:2006xs}.

Among their generation mechanisms, there have been proposed %cosmological MF generation mechanisms
processes related to phase transitions in the early Universe~\cite{1988PhRvD..37.2743T,1989ApJ...344L..49Q}, primordial scalar~\cite{Ichiki:2006cd,Naoz:2013wla,Flitter:2023xql} and vector perturbations~\cite{Banerjee:2003xk,Durrer:2006pc} as well as astrophysical ones seeding battery-induced MFs~\cite{1950ZNatA...5...65B}. In particular, in the last years there has been witnessed a rekindled interest connecting the origin of primordial MFs with primordial black holes (PBHs)~\cite{Safarzadeh:2017mdy,Araya:2020tds,Papanikolaou:2023nkx}. As it was recently shown in~\cite{Papanikolaou:2023nkx} supermassive PBHs furnished with a disk can generate %à la Biermann 
through the Biermann-battery mechanism the seed for the primordial MFs of $10^{-18}\mathrm{G}$ observed in intergalactic scales~\cite{2011ApJ...733L..21D}.

Interestingly enough, PBHs, firstly introduced in '70s can address a swathe of modern cosmological enigmas. In particular, they can naturally account for a fraction or even the totality of dark matter~\cite{Chapline:1975ojl,Clesse:2017bsw} explaining as well LSS formation through Poisson fluctuations~\cite{Meszaros:1975ef,Afshordi:2003zb,1984MNRAS.206..315C} and providing the seeds of the supermassive black holes residing in galactic centres~\cite{ Bean:2002kx,DeLuca:2022bjs,Li:2023zyc,Su:2023jno}. At the same time, they are associated with numerous GW %gravitational-wave (GW) KNG We have introduced the shortcut GW earlier
signals related to PBH merging events~\cite{Nakamura:1997sm, 
Eroshenko:2016hmn,Raidal:2017mfl},  Hawking radiation~\cite{Anantua:2008am,Dong:2015yjs,Ireland:2023avg} and enhanced cosmological~\cite{Saito_2009,Bugaev:2009zh,Kohri:2018awv} adiabatic and isocurvature perturbations~\cite{Papanikolaou:2020qtd,Domenech:2020ssp,Domenech:2021and,Papanikolaou:2022chm}. See here for recent reviews~\cite{Sasaki:2018dmp,Domenech:2021ztg}.

In this work, we derive 
the GW signal induced by the magnetic anisortropic stress of a population of magnetised supermassive PBHs. 
Accounting as well for GW constraints from BBN we are able to set tight constraints on the abundances of supermassive PBHs promoting in this way the magnetically induced GWs (MIGWs) as a novel portal shedding light to the field of PBH physics.

%%%%%%%%%%%%%%%%%%%%%%%%%%% SECTION 2: The seed primordial magnetic field à la Biermann. 
%%%%%%%%%%%%%%%%%%%%%%%%%%%%%%%%%%%

\section{The seed primordial magnetic field à la Biermann}
PBH accretion disks have been recently proposed as a candidate to generate the seed primordial MFs threading the intergalactic medium~\cite{Safarzadeh:2017mdy,Araya:2020tds,Papanikolaou:2023nkx}. In particular, the ab initio generation of a seed MF requires the relative motion between negative and positive charges, conditions which can be %easily 
achieved in a highly turbulent medium such as the primordial plasma between BBN and recombination ~\cite{1974SvA....18..157K,1974SvA....18..300K,Trivedi:2018ejz,RoperPol:2021gjc}. Under such conditions, a Biermann battery mechanism operates whenever the energy density and temperature gradients are not parallel to each other \cite{1983MNRAS.204.1025B}. Consequently, one is inevitably met with the following magnetic field induction equation 
\begin{eqnarray}
    \partial_t {\bf B}= \nabla \left({\bf u} \times {\bf B}\right) -\frac{c k_B}{e} \frac{\nabla \rho \times \nabla T}{\rho}\,, 
    \label{eq:induction}
\end{eqnarray}
where the second term in the right hand side is the Biermann battery one.

In order to derive the Biermann battery induced MF one should assume an equation of state (EoS) for the vortex-like moving plasma around the black hole. Doing so, we assume a locally isothermal disk around the PBH~\cite{2010ApJ...724..730D}, where the density and the pressure are related through the following relation:
\begin{eqnarray}\label{eq:local_isothermal}
    p(R,\phi,z)= \rho(R, \phi, z) c_s^2(R)\,,
\end{eqnarray}
where $(R, \phi, z)$ are the cylindrical coordinates. This EoS can describe quite well a gas that radiates internal energy gained by shocks~\cite{1999ApJ...526.1001L}, here produced by the turbulent motion of the primordial plasma expected after BBN and before recombination era~\cite{1974SvA....18..157K,1974SvA....18..300K,Trivedi:2018ejz,RoperPol:2021gjc}. 

At the end, accounting for the random spatial distribution of PBHs and considering monochromatic PBH mass distributions, after a long but straightforward calculation [For more details see~\cite{Papanikolaou:2023nkx}] one can extract the MF power spectrum, which can be recast as~\footnote{To extract \Eq{eq:P_B} we followed the prescription described in the Appendix of~\cite{Papanikolaou:2023nkx} considering that that PBH mass is of the order of the mass within the cosmological horizon at the time of PBH formation~\cite{Musco_2005}.}
\beq\label{eq:P_B}
\begin{split}
P_B(k,t_\mathrm{s}) & \simeq 4\times 10^{-86}q^2\ell^4_R\Omega^2_\mathrm{PBH,f} \\ & \times \left(\frac{M}{10^{10}M_\odot}\right)^2\left(\frac{k}{\mathrm{Mpc}^{-1}}\right)^3 \left[\mathrm{G^2Mpc^3}\right],
\end{split}
\eeq
where $\ell_R=R_\mathrm{d}/R_\mathrm{ISCO}$ is the ratio of the radius of the disk, $R_\mathrm{d}$ over the radius of the innermost stable circular orbit, $R_\mathrm{ISCO}$ and $q=H_\mathrm{d}/R_\mathrm{ISCO}$ is the ratio of the thickness of the disk $H_\mathrm{d}$ over $R_\mathrm{ISCO}$ which is less than $1$ since \Eq{Eq:B_intergalactic} was extracted within the thin-disk limit where one is usually met with sub-Eddington accretion~\cite{1973A&A....24..337S,1972A&A....21....1P,
1974MNRAS.168..603L}. It is important to notice that the above mentioned MF power spectrum was extracted at saturation time $t_\mathrm{s}$, namely at the end of the linear growth phase of the MF, [See the Biermann-battery term in \Eq{eq:induction}] and for scales larger than the PBH mean separation scale, so as not to enter the non-linear regime. This imposes a UV-cutoff scale $k_\mathrm{UV}$ which can be recast straightforwardly as $k_\mathrm{UV} = 10^{19}M_\odot\Omega^{1/3}_\mathrm{PBH,f}\mathrm{Mpc}^{-1}/M$~\cite{Papanikolaou:2023nkx}.

One can derive as well
the mean MF amplitude which is defined as 
\beq\label{eq:mean_B}
\langle |\boldmathsymbol{B}_\boldmathsymbol{k}|\rangle  \equiv \sqrt{\frac{k^3 P_B(k)}{2\pi^2}}.
\eeq
Accounting therefore for cosmic expansion, i.e. $B\sim a^{-2}$, and plugging now \Eq{eq:P_B} into \Eq{eq:mean_B} one gets the mean MF amplitude in intergalactic scales, i.e. $k\sim 100\mathrm{Mpc}^{-1}$, which reads as
\beq\label{Eq:B_intergalactic}
B\sim 10^{-30}q\left(\frac{\ell_\mathrm{R}}{10^6}\right)^2\left(\frac{M_\mathrm{PBH}}{10^{14}M_\odot}\right)^{5/2}\quad (\mathrm{G}).
\eeq
Interestingly enough, by taking typical values of $q\sim 0.001 - 1$ and varying the parameter $\ell_R$ with the range $\ell_R\in[10^{2},10^{11}]$ depending on the accretion rate~\cite{McKinney:2012vh}, one can produce for PBH masses $M\in[10^{10},10^{16}]M_\odot$~\footnote{We need to point out here that in order to generate through the Biermann battery mechanism the seed primordial MF which is necessary to give rise to a MF amplitude of the order $10^{-18}\mathrm{G}$, threading the intergalactic medium, one needs to consider PBH masses higher than $10^{10}M_\odot$ as it was shown in~\cite{Papanikolaou:2023nkx}.} a seed primordial MF of the order $10^{-32}-10^{-29}\mathrm{G}$, which is actually the minimum seed MF amplitude needed to give rise, through dynamo/turbulent amplification, to the present-day average magnetic field of order $10^{-18}\mathrm{G}$ in intergalactic scales~\cite{Vachaspati:2020blt}.

%%%%%%%%%%%%%%%%%%%%%%%%%%% SECTION 3: The magnetic field anisotropic stress.
%%%%%%%%%%%%%%%%%%%%%%%%%%%%%%%%%%%
\section{The magnetic field anisotropic stress} Let us now extract the magnetic anisotropic stress induced by such a MF power spectrum. In particular, regarding the stress-energy tensor associated to a magnetic field $\boldmathsymbol{B}$, this can be recast in the following covariant form:
\beq\label{eq:magnetic_T_mu_nu}
T^{(B)}_{ij} \equiv\frac{1}{4\pi}\left[\frac{B^2 g_{ij}}{2} - B_iB_j\right].
\eeq
From \Eq{eq:magnetic_T_mu_nu} one can define an associated anisotropic stress as follows
\beq\label{eq:magnetic_anisotropic_stress}
\Pi_{ij}(\boldmathsymbol{k}) \equiv \left(P^l_{i}P^m_{j} - \frac{P_{ij}P^{lm}}{2}\right)T_{lm}(\boldmathsymbol{k}),
\eeq
where $P_{ij}$ is a projection matrix defined as $P_{ij}\equiv \delta_{ij} - \hat{\boldmathsymbol{k}}_i\hat{\boldmathsymbol{k}}_j$ and $\hat{\boldmathsymbol{k}} = \boldmathsymbol{k}/k$. From \Eq{eq:magnetic_anisotropic_stress} one can define the equal-time 2-point correlator of the magnetic anisotropic stress as:
\beq
\langle \Pi_{ij}(\boldmathsymbol{k},\eta)\Pi_{ij}(\boldmathsymbol{q},\eta)\rangle    \equiv \Pi_B(k,\eta)\delta(\boldmathsymbol{k},\boldmathsymbol{q}),
\eeq
where $\Pi_B(\boldmathsymbol{k},\eta)$ is the power spectrum of the magnetic anisotropic stress related with the magnetic field power spectrum $P_B(k,\eta)$ as follows~\cite{Caprini:2006jb}:
\beq\label{eq:Pi_B_vs_P_B}
\Pi_B(k,\eta) = \int \mathrm{d}^3\boldmathsymbol{q} P_B(q,\eta)P_B(|\boldmathsymbol{q}-\boldmathsymbol{k}|,\eta) (1+\gamma^2) (1+\beta^2),
\eeq
where $\gamma = \hat{\boldmathsymbol{k}}\cdot \hat{\boldmathsymbol{q}}$ and $\beta = \hat{\boldmathsymbol{k}}\cdot \widehat{\boldmathsymbol{k} - \boldmathsymbol{q}}$. 

Introducing now the auxiliary variables $v$ and $u$ such as that $u=|\boldmathsymbol{q}-\boldmathsymbol{k}|/k$ and $v=q/k$ after some algebraic manipulations one can recast the above equation in the following form:
\beq
\label{eq:Pi_B_vs_P_B_u_v}
\begin{split}
& \Pi_B(k,\eta) = 2\pi k^3 \int_{0}^\mathrm{\frac{k_\mathrm{UV}}{k}} \mathrm{d}v \int_{|1-v|}^{1+v} \mathrm{d} u P_B(kv,\eta)P_B(ku,\eta) uv \\  & \times
\left[1+\frac{(1+v^2-u^2)^2}{4v^2}\right]\left[1+ \left(1 - \frac{1 + v^2 - u^2}{2v}\right)^2\right].
\end{split}
\eeq

At the end, plugging \Eq{eq:P_B} in \Eq{eq:Pi_B_vs_P_B} one gets that

\beq\label{Pi_B_Biermann}
\begin{split}
\Pi_B(k,\eta) & \simeq 10^{-170} q^4\ell^8_R \Omega^4_\mathrm{PBH,f}10^{-174}\left(\frac{k}{\mathrm{Mpc}^{-1}}\right)^9 \\ & \times \left(\frac{M}{10^{10}M_\odot}\right)^4 f\left(\frac{k_\mathrm{UV}}{k}\right)\mathrm{G^4 Mpc^{3}},
\end{split}
\eeq
where $f(\frac{k_\mathrm{UV}}{k})$ is the double integral over $u$ and $v$ which is defined as follows:
\beq
\begin{split}
f\left(\frac{k_\mathrm{UV}}{k}\right) \equiv  & \int_{0}^{\frac{k_\mathrm{UV}}{k}}\mathrm{d} v \int_{|1-v|}^{1+v}\mathrm{d} u u^4v^4 \left[1+\frac{(1+v^2-u^2)^2}{4v^2}\right] \\ & \times \left[1+ \left(1 - \frac{1 + v^2 - u^2}{2v}\right)^2\right].
\end{split}
\eeq
After performing the integration one can show that
\beq
\begin{split}
& f\left(y\equiv \frac{k_\mathrm{UV}}{k}\right) = \frac{32}{45}y^9 - \frac{131 y^8}{60} + \frac{293 y^7}{98} - \frac{3323 y^6}{1680} \\ & + \frac{27229 y^5}{50400} - \frac{2 y^4}{105} + \frac{26 y^3}{3465} - \frac{2 y^2}{1155} + \frac{
 8 y}{15015} - \frac{19711}{6306300}.
 \end{split}
\eeq
In the region away from the UV-cutoff scale $k_\mathrm{UV}$, namely when $k_\mathrm{UV}/k\gg 1$, one obtains that $f\left(y\equiv \frac{k_\mathrm{UV}}{k}\right)\simeq (32/45)(k_\mathrm{UV}/k)^9$.

%%%%%%%%%%%%%%%%%%%%%%%%%%% SECTION 4: Magnetically induced gravitational waves.
%%%%%%%%%%%%%%%%%%%%%%%%%%%%%%%%%%%

\section{Magnetically induced gravitational waves}
Having extracted above the power spectrum of the magnetic anisotropic stress we study here the dynamics of the tensor perturbations $h_\boldmathsymbol{k}$ induced by such an anisotropic stress. In particular, the equation of motion for $h_\boldmathsymbol{k}$ can be recast as~\cite{Caprini:2006jb}:
\beq\label{eq:tensor_eq_motion}
h_\boldmathsymbol{k}^{s,\prime\prime} + 2\mathcal{H}h_\boldmathsymbol{k}^{s,\prime} + k^{2}h^s_\boldmathsymbol{k} = \frac{8\pi G}{a^2}\sqrt{\Pi_B(k,\eta)},
\eeq 
where $s = (+), (\times)$ stands for the two polarisation states of the tensor modes in general relativity. 

An analytic solution to \Eq{eq:tensor_eq_motion} can be obtained by using the Green's function formalism. Namely, one gets that
\bea
\label{tensor mode function}
h^s_\boldmathsymbol{k} (\eta)  = 8\pi G \int^{\eta}_{\eta_0}\mathrm{d}\bar{\eta}\,  g_\boldmathsymbol{k}(\eta,\bar{\eta})\frac{\sqrt{\Pi_B(k,\bar{\eta})}}{a^2(\bar{\eta})},
\eea
where $g_\boldmathsymbol{k}(\eta,\bar{\eta})$ is the Green's function being the solution of the homogeneous equation \eqref{eq:tensor_eq_motion} without the source term [See~\cite{Caprini:2006jb} for more details]. For the case of an RD era, $w=1/3$, when PBHs typically form, one gets that 
\beq\label{eq:G_k_RD}
kg^{\mathrm{RD}}_\boldmathsymbol{k}(\eta,\bar{\eta})  = \sin(x - \bar{x}).
\eeq

One then can define the GW spectral abundance as $\OmegaGW (\eta,k)\equiv \frac{1}{\rho_\mathrm{c}}\frac{\mathrm{d}\rhoGW}{\mathrm{d}\ln k}$, where $\rho_\mathrm{c}=3H^2/(8\pi G)$ is the critical energy density, and show that it can be recast as~\cite{Maggiore:1999vm,Kohri:2018awv} 
\bea
\label{Omega_GW_RD}
\OmegaGW (\eta,k) = \frac{1}{24}\left[\frac{k}{\calH(\eta)}\right]^{2}\overline{\mathcal{P}}_{h}(\eta,k), 
\eea
where $\calH$ is the conformal Hubble parameter and $\mathcal{P}_{h}(\eta,k),$ is the tensor power spectrum defined as 
\beq\label{eq:P_h}
\mathcal{P}_{h}(\eta,k) \equiv \frac{k^3|h_\boldmathsymbol{k}|^2}{2\pi^2}.
\eeq
The bar denotes averaging over the sub-horizon oscillations of the tensor field, which is done in order to extract the envelope of the GW spectrum at those scales.

Combining therefore \Eq{Pi_B_Biermann}, \Eq{tensor mode function}, \Eq{eq:G_k_RD}, \Eq{eq:P_h} and \Eq{Omega_GW_RD}, one obtains that the GW spectral abundance will read as
\beq
\begin{split}
\OmegaGW (\eta,k) & = 10^{-70}I^2(x)\left(\frac{k}{\mathrm{Mpc}^{-1}}\right) \\ & \times \left(\frac{10^{10}M_\odot}{M}\right)^{4}\ell^8_R\Omega^7_\mathrm{PBH,f},
\end{split}
\eeq
where $x=k\eta$ and $I(x)$ is defined as
\beq
I(x) \equiv \int_{x_\mathrm{dyn}}^{x}\mathrm{d}\bar{x}\frac{\sin(x-\bar{x})}{\bar{x}^2}\left[1-\frac{x^2_\mathrm{dyn}}{\bar{x}^2}\right]^2,
\eeq
where $\eta_\mathrm{dyn}$ stands for the conformal disk dynamical time and $x_\mathrm{dyn} = k\eta_\mathrm{dyn} \simeq 1$ since the disk establishes very quickly its hydrostatic equilibrium on the vertical axis soon after the PBH formation time, which is standardly considered as the time when the typical size of the collapsing overdensity region $r\sim 1/k$ crosses the cosmological horizon, i.e. when $k=aH$. Thus, one has that $\eta_\mathrm{dyn} \simeq \eta_\mathrm{f}$ and that $x_\mathrm{dyn} = k\eta_\mathrm{dyn} \simeq k\eta_\mathrm{f} = k/(a_\mathrm{f}H_\mathrm{f}) = 1 $. 
%%%%%%%%%%%%%%%%%%%%%%%%%%% SECTION 5: Non-Amplified magnetically induced gravitational waves.
%%%%%%%%%%%%%%%%%%%%%%%%%%%%%%%%%%%

\section{Non-amplified magnetically induced gravitational waves}
In what follows, we will account for the contribution of the magnetic anisotropic stress during the time interval where the Biermann battery mechanism operates within its linear growth regime up to $t=t_\mathrm{s}$, hence underestimating the GW signal and considering that the tensor modes propagate as free GWs up to our present day soon after the end of the era of the linear growth of $\boldmathsymbol{B}$ at $t=t_\mathrm{s}$.  

Thus, for $x=x_\mathrm{s}=\sqrt{2}x_\mathrm{dyn}$~\cite{Papanikolaou:2023nkx} one can check numerically that $I^2(x_\mathrm{s})\simeq 10^{-5}$. At the end, accounting for the fact that $a^2H^2\propto a^{-2}$ in the RD era one gets that 
\beq\label{eq:OmegaGW_t_s}
\begin{split}
\OmegaGW (\eta_\mathrm{s},k) & = 10^{-75}\left(\frac{k}{\mathrm{Mpc}^{-1}}\right) \\ & \times \left(\frac{10^{10}M_\odot}{M}\right)^{4}\ell^8_R\Omega^7_\mathrm{PBH,f}.
\end{split}
\eeq

One then can compute the GW spectral abundance $\OmegaGW(\eta,k)$ at our present epoch as follows:
\beq
\begin{split}
\OmegaGW(\eta_0,k) & = \frac{\rhoGW(\eta_0,k)}{\rho_\mathrm{c}(\eta_0)} = \frac{\rhoGW(\eta_\mathrm{s},k)}{\rho_\mathrm{c}(\eta_\mathrm{s})}\left(\frac{a_\mathrm{s}}{a_\mathrm{0}}\right)^4 \frac{\rho_\mathrm{c}(\eta_\mathrm{s})}{\rho_\mathrm{c}(\eta_0)} 
\\ & = c_g \Omega^{(0)}_\mathrm{r} \OmegaGW(\eta_\mathrm{s},k),
\end{split}
\eeq
where $c_g=\frac{\rho_\mathrm{r,s}a^4_\mathrm{s}}{\rho_\mathrm{r,0}a^4_0}\simeq 0.4$~\cite{Espinosa:2018eve}, $\Omega^{(0)}_\mathrm{r}\sim 10^{-5}$ and we have taken into account as well that $\Omega_\mathrm{GW}\sim a^{-4}$. The index $0$ refers to the present time. Finally, $\OmegaGW(\eta_0,k)$ will be recast as
\beq\label{Omega_GW_RD_0}
\begin{split}
\Omega_\mathrm{GW}(\eta_0,k) & 
\simeq 3\times 10^{-81}\left(\frac{k}{\mathrm{Mpc}^{-1}}\right)\left(\frac{10^{10}M_\odot}{M}\right)^{4}q^4\ell^8_R\Omega^7_\mathrm{PBH,f}.
\end{split}
\eeq

Regarding now the GW frequency this will be given by $f=k/(2\pi a_0)$ where we conventionally take $a_0=1$. Thus, since $k\leq k_\mathrm{UV}$, one can extract an upper bound constraint on the GW frequency, namely that 
\beq
\begin{split}
f & \leq f_\mathrm{max} =\frac{k_\mathrm{UV}}{2\pi}= 10^5\frac{M_\odot}{M}\Omega^{1/3}_\mathrm{PBH,f} \\ & \leq 3\times 10^{-7}\left(\frac{10^{10}M_\odot}{M}\right)^{5/6}\leq 3 \times 10^{-7}\quad \mathrm{(Hz)} ,
\end{split}
\eeq
since $M>10^{10}M_\odot$ and $\Omega_\mathrm{PBH,f}<2.6\times 10^{-5}\sqrt{\frac{M}{10^{10}M_\odot}}$ so as not to overproduce PBHs at matter-radiation equality. Therefore the relevant MIGW signal is far away from the frequency bands of the Laser Space Inferometer Antenna (LISA)~\cite{Caprini:2015zlo,Karnesis:2022vdp}, Einstein Telscope (ET)~\cite{Maggiore:2019uih} and Big Bang Observer (BBO)~\cite{Harry:2006fi} GW detectors. Potentially, it can be well within the Square Kilometer Array (SKA)~\cite{Janssen:2014dka}, the NANOGrav~\cite{NANOGrav:2023gor} and the PTA~\cite{Reardon:2023gzh,Antoniadis:2023rey} frequency detection band. However, for smaller mass PBHs furnished with a disk, which however will not be able to seed primordial MFs~\cite{Papanikolaou:2023nkx} the GW frequency will increase and can be well within LISA, ET and BBO frequency detection bands.

Apart from the GW frequency, one should check as well the GW amplitude \eqref{Omega_GW_RD_0}. In particular, considering that $k\leq k_\mathrm{UV}$, $\ell_R\leq 10^{11}$ and accounting for \Eq{Omega_GW_RD_0} as well as for the fact that for $M>10^{10}M_\odot$, $\Omega_\mathrm{PBH,f}<2.6\times 10^{-5}\left(\frac{M}{10^{10}M_\odot}\right)^{1/2}$, one gets an upper bound on $\Omega_\mathrm{GW}(\eta_0)$ which reads as
\beq
\Omega_\mathrm{GW}(\eta_0)<7\times 10^{-18}\left(\frac{10^{10}M_\odot}{M}\right)^{4/3}\leq  7 \times 10^{-18},
\eeq
and which is slightly below the lowest GW sensitivity of current and future GW detectors being of the order of $10^{-15}$. Consequently, these magnetically induced GWs can hardly be detected by current/future GW detectors.

%%%%%%%%%%%%%%%%%%% SECTION 6: Amplified magnetically induced gravitational waves. %%%%%%%%%%%%%%%%%%%%%%%
\section{Amplified magnetically induced gravitational waves}
However, up to now, we have not accounted for the turbulent and galactic dynamo or the magnetorotational instability \cite{velikhov1959stability,1960PNAS...46..253C,1991ApJ...376..214B} amplification which will play a significant role after matter-radiation equality during the non-linear growth of the matter perturbations. These amplification mechanisms can significantly enhance the MF amplitude and at the end the GW signal potentially putting it within the sensitivity bands of current/future GW detectors. To account therefore for these effects we introduce the amplification factor $\alpha(k)$ as the ratio between the amplified MF over the non-amplified MF, i.e.
\beq
\alpha (k) = \frac{B^\mathrm{amplified}(k)}{B^\mathrm{non-amplified}(k)}.
\eeq

To extract the function $\alpha(k)$ over the different scales $k$ at hand one should run high-cost numerical MHD simulations so as to account for the various turbulent/galactic dynamo and instability processes, something which is beyond the scope of this work. Thus, in order to make below quantitative predictions for the GW signal, we will assume an effective power-law toy model for $\alpha(k)$ which can be recast as
\beq\label{eq:alpha_def}
\alpha(k) = \alpha(k_*)\left(\frac{k}{k_*}\right)^{n_B},
\eeq
where $k_*$ is a pivot scale and $n_B$ is the amplification spectral index which should be greater or equal to zero since in small scales one expects to have a greater MF amplification~\footnote{Concerning the order of magnitude for the amplitude of the MFs in the Universe, there is strong evidence for a pre-galactic seed magnetic field of the order of $10^{-16}-10^{-18}\mathrm{G}$~\cite{2010Sci...328...73N,2011ApJ...733L..21D} on intergalactic scales while on galactic scales we observe MFs with present day amplitudes up to $10^{-7}\mathrm{G}$~\cite{2004NewAR..48..763V,2011ApJ...728...97V,2019A&A...622A..16O}. In smaller scales, the MF intensity is strongly influenced by the presence of interstellar gas and the proximity to stars. For instance in the vicinity of the Earth the interplanetary magnetic field is $10^{-4}$G \cite{2013LRSP...10....5O}.}. In what follows, we will consider as our pivot scale the characteristic intergalactic scale $k_*=100\mathrm{Mpc}^{-1}$ where we know from observations that $B\sim 10^{-18}\mathrm{G}$~\cite{2011ApJ...733L..21D} and thus $\alpha(k_*)$ will read as
\beq\label{eq:alpha_star}
\begin{split}
& \alpha(k_*=100\mathrm{Mpc}^{-1})  =  \frac{10^{-18}}{10^{-30}\mathrm{G}q\left(\frac{\ell_\mathrm{R}}{10^6}\right)^2\left(\frac{M_\mathrm{PBH}}{10^{14}M_\odot}\right)^{5/2}} \\ &  = \frac{10^{22}}{q}\left(\frac{10^6}{\ell_R}\right)^2\left(\frac{10^{10}M_\odot}{M}\right)^{5/2},
\end{split}
\eeq
where we have used \Eq{Eq:B_intergalactic} for the non-amplified MF amplitude on intergalactic scales.

This amplification effect will give an extra $a^2(k)$ factor at the level of the MF power spectrum $P_B(k)$ since $P_B(k)\propto B^2_k$. At the end, one is met with a rough overall $a^4(k)$ amplification factor at the level of the GW signal since $\OmegaGW \propto \int \int P^2_B$ as we can see from \Eq{tensor mode function}, \Eq{eq:Pi_B_vs_P_B_u_v} and \Eq{Omega_GW_RD}. Thus, multiplying \Eq{Omega_GW_RD_0} with $\alpha^4(k)$ one gets that 

\beq\label{eq:Omega_GW_amplified}
\begin{split}
\Omega_\mathrm{GW}(k,\eta_0) & \simeq 3\times 10^{55-8n_B}\left(\frac{k}{\mathrm{Mpc}^{-1}}\right)^{4n_B+1}\\ & \times\left(\frac{10^{10}M_\odot}{M}\right)^{14} \Omega^7_\mathrm{PBH,f}
\end{split}
\eeq

\begin{figure}[t!]
\begin{center}
\includegraphics[width=0.495\textwidth]{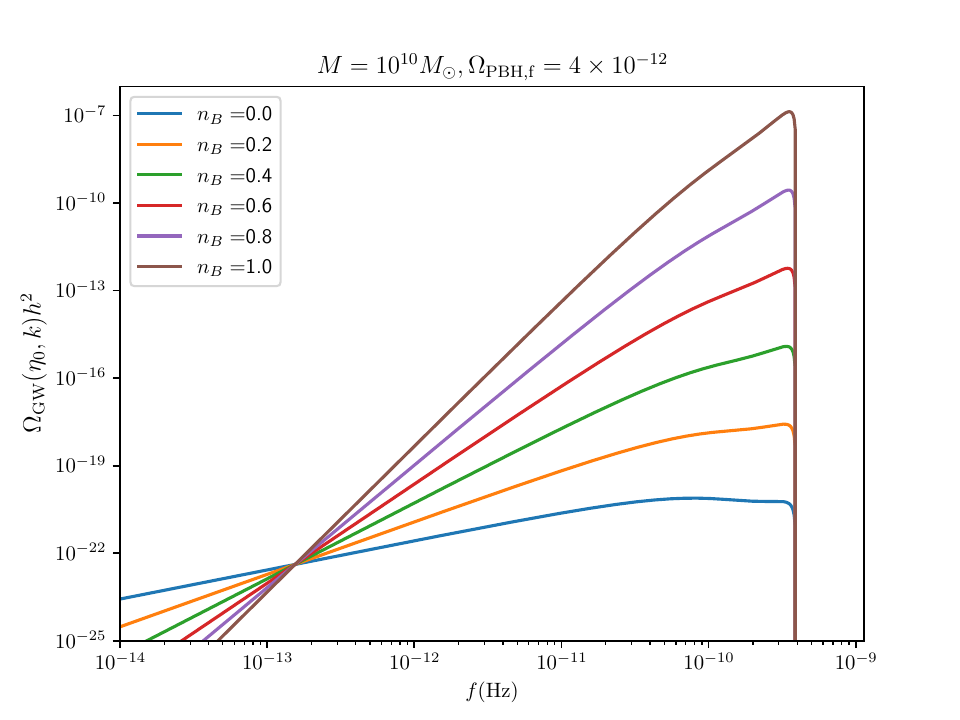}
\caption{{\it{The gravitational-wave spectrum induced by a population of magnetised PBHs with mass $M=10^{10}M_\odot$ and initial PBH abundance at formation time $\Omega_\mathrm{PBH,f}=4\times 10^{-12}$ for different values of the amplification spectral index $n_B$.}}}
\label{fig:Omega_GW_vs_nB}
\end{center}
\end{figure}
%

%%%%%%%%%%%%%%%%%%% SECTION 7: Constraints on supermassive primordial black holes. %%%%%%%%%%%%%%%%%%%%%%%
\section{Constraints on supermassive primordial black holes}
Let us see now how one can constrain the abundances of such supermassive PBHs using the aforementioned GW portal. Interestingly enough, one can set an upper bound on $\Omega_\mathrm{PBH,f}$ by accounting for the contribution of the GWs to the effective number of extra neutrino species $\Delta N_\mathrm{eff}$. In particular, one can translate the upper bound from Planck for the amplitude of GWs today, namely that $\Omega_\mathrm{GW}(\eta_0)\leq 10^{-6}$~\cite{Caprini:2018mtu,Planck:2018vyg}, to an upper bound on $\Omega_\mathrm{PBH,f}$.

As we see from \Eq{eq:Omega_GW_amplified}, for $k\geq 100\mathrm{Mpc}^{-1}$ the GW amplitude increases with $n_B$. Thus, in order to get a conservative constraint on $\Omega_\mathrm{PBH,f}$ we will choose the flat case where $n_B=0$. At the end, by using \Eq{eq:Omega_GW_amplified} for $n_B=0$ and setting $k=k_\mathrm{UV}$, since at $k\sim k_\mathrm{UV}$ one gets the maximum GW amplitude [See \Fig{fig:Omega_GW_vs_nB}], we obtain a conservative upper bound constraint on $\Omega_\mathrm{PBH,f}$ by just simply requiring that $\Omega_\mathrm{GW}(\eta_0)\leq 10^{-6}$ reading as~\footnote{It is important to highlight here that the constraint \eqref{eq:Omega_PBH_f_bound} on the PBH abundances is valid only for PBH masses higher than $10^{10}M_\odot$. This is because our pivot amplification factor $\alpha(k_*)$ was computed at the intergalactic scale $k_\mathrm{intg}=100\mathrm{Mpc}^{-1}$ assuming that our Biermann-battery mechanism can give rise to present day intergalactic MFs of the order of $10^{-18}\mathrm{G}$~\cite{Papanikolaou:2022hkg}. If now we operate the proposed Biermman-battery mechanism with smaller mass PBHs, not being able to give rise to the present day intergalactic MFs as it was shown in~\cite{Papanikolaou:2023nkx}, we will not be able to have an estimate on the pivot amplification factor and at the end on the present day MIGW signal.}
\beq\label{eq:Omega_PBH_f_bound}
\Omega_\mathrm{PBH,f}\leq 2.5\times 10^{-10}\left(\frac{M}{10^{10}M_\odot}\right)^{45/22}.
\eeq

\begin{figure}[h!]
\begin{center}
\includegraphics[width=0.495\textwidth]{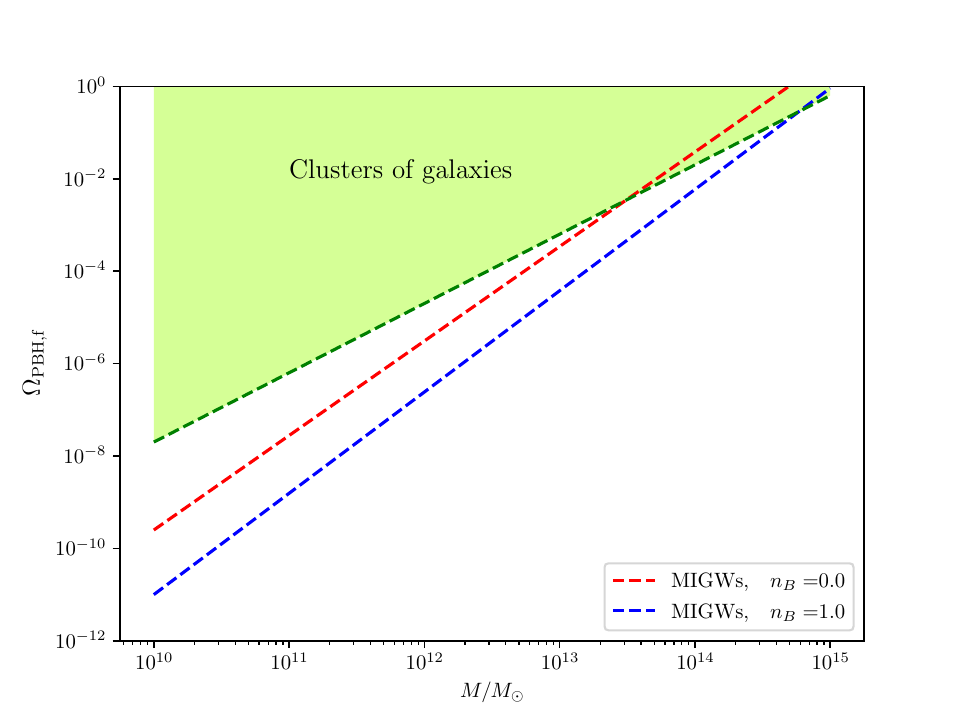}
\caption{{\it{The constraints on the initial PBH abundance at formation time $\Omega_\mathrm{PBH,f}$ as a function of the PBH mass $M$ for $n_B=0$ (dashed red curve) and $n_B=1$ (dashed blue curve). In the green region we show the constraints on $\Omega_\mathrm{PBH,f}$ from galaxy clusters~\cite{Carr:2018rid}. }}}
\label{fig:Omega_PBH_f_vs_M_constraints}
\end{center}
\end{figure}

Remarkably, this upper bound constraint on $\Omega_\mathrm{PBH,f}$ is comparable and in some mass regions tighter compared to constraints on $\Omega_\mathrm{PBH,f}$ from LSS probes, which were derived by simply requiring that galaxies should not form too early~\cite{Carr:2018rid,Carr:2020gox}. Interestingly, if one increases the amplification spectral index $n_B$ they get tighter constraints on $\Omega_\mathrm{PBH,f}$ up to $10^{15}M_\odot$. See \Fig{fig:Omega_PBH_f_vs_M_constraints} for more details. Therefore, the portal of GWs induced by magnetised PBHs is inevitably promoted as a new probe to explore the enigmatic nature of supermassive PBHs.

Here, it is important to highlight however that we did not consider $\mu$ and $y$ distortions of the Cosmic Microwave Background (CMB) affected by PBH formation which put strong constraints on $\Omega_\mathrm{PBH,f}$ assuming Gaussian primordial perturbations~\cite{Kohri:2014lza,Karam:2022nym}. These strong constraints can in general be evaded assuming non-Gaussian cosmological perturbations~\cite{Nakama:2016kfq,Kawasaki:2019iis,Hooper:2023nnl}, hence we do not consider them in this work.
%

%%%%%%%%%%%%%%%%%%% SECTION 8: Discussion. %%%%%%%%%%%%%%%%%%%%%%%

\section{Discussion}
The origin of cosmic magnetic fields constitutes one of the longstanding issues in cosmology. Among their generation mechanisms the portal of magnetised PBHs seeding battery-induced cosmic MFs seems one of the most promising. 

In this work, we have considered a population of supermassive PBHs furnished with a locally isothermal disk which can generate through Biermann-battery a seed primordial MF in the intergalactic scales. In particular, we derived for the first time %to the best of our knowledge 
the GWs induced by the magnetic anisotropic stress of such a population of magnetised PBHs. 

Interestingly enough, by accounting for the contribution of the MIGWs to the effective number of extra neutrino
species $\Delta N_\mathrm{eff}$ and adopting an effective model for the galactic/turbulent dynamo amplification of the magnetic field, we set upper bound constraints on the abundances of supermassive PBHs at formation time, $\Omega_\mathrm{PBH,f}$ as a function of the their masses which reads as
\beq
\Omega_\mathrm{PBH,f}\leq 2.5\times 10^{-10}\left(\frac{M}{10^{10}M_\odot}\right)^{45/22}.
\eeq

In particular, as one may see from \Fig{fig:Omega_PBH_f_vs_M_constraints}, these constraints are comparable and in some mass ranges even tighter compared to constraints on $\Omega_\mathrm{PBH,f}$ derived from clusters of galaxies. One should account as well for constraints on supermassive PBHs due to dark matter halo accretion after matter-radiation equality as discussed in~\cite{Serpico:2020ehh}, where it was derived a mass-independent upper bound constraint on the contribution of PBHs to dark matter, $f_\mathrm{PBH}\equiv \Omega_\mathrm{PBH}/\Omega_\mathrm{DM}$, of the order of $3\times 10^{-9}$. However, the aforementioned accretion constraint is not so robust for PBH masses larger than $10^{4}M_\odot$, as the ones considered here, since one finds super-Eddington accretion at this high mass range.

At this point, it is noteworthy to mention that in the present work we consider thin accretion disks usually exhibiting sub-Eddington accretion~\cite{1973A&A....24..337S,1972A&A....21....1P,
1974MNRAS.168..603L}, which in our case operate only during the linear growth of the magnetic field up to few dynamical times~\cite{Papanikolaou:2023nkx}. Note also that the only place in our analysis where one is met with a dependence on the accretion model is the dimensionless parameter $\ell_R$ giving us the radial size of the disk, which in general depends on the accretion rate~\cite{McKinney:2012vh}. Interestingly enough, this parameter cancels out in the final expression \Eq{eq:Omega_GW_amplified} for the GW signal today since one needs to multiply $\alpha^4\propto \ell^{-4}_R$~[See \Eq{eq:alpha_def} and \Eq{eq:alpha_star}] with \Eq{Omega_GW_RD_0}. In order to extract a potentially more stringent accretion constraint on the PBH abundances in the mass region $M_\mathrm{PBH}>10^{4}M_\odot$ one needs in principle to run dedicated hydro-dynamical simulations in a cosmological setting, going beyond the scope of the current work. Thus, in the absence of numerical simulations for accretion in the very high PBH mass regime~\cite{Serpico:2020ehh}, it can be claimed that the portal of MIGWs can act as a novel alternative probe to constrain the abundances of supermassive PBHs.

This portal of MIGWs can also be used in order to constrain lower mass PBHs furnished with Biermann battery induced MF which however do not generate the necessary seed primordial MF to give rise to a MF amplitude of the order $10^{-18}\mathrm{G}$, threading the intergalactic medium~\cite{Papanikolaou:2023nkx}. Nevertheless, there exist other MF generation mechanisms, like the cosmic battery one~\cite{1998ApJ...508..859C,2006ApJ...652.1451C,2018MNRAS.473..721C}, able to give a very high MF amplification on intergalactic scales operating on lower black hole masses, i.e. $M<10^{10}M_\odot$.

Furthermore, it is important to emphasize here that within this work we assume the standard PBH formation scenario where PBHs form out of the collapse of enhanced cosmological perturbations with a mass of the order of that within the cosmological horizon~\cite{Musco:2012au} at the time of  PBH formation, remaining agnostic on the specific cosmological model giving rise to enhanced cosmological perturbations. In order to access the exact PBH mass distribution, one should choose a particular cosmological model giving rise to PBH formation and account for the critical collapse scaling law for the PBH mass spectrum~\cite{Niemeyer:1997mt, Musco:2008hv} as well for the effect of primordial non-Gaussianities which are necessary to avoid the $\mu$ and $y$ distortion constraints. These effects lead in principle to extended PBH mass functions. In this work, we assume for simplicity a monochromatic PBH mass distribution. This choice can be sufficiently justified only for sharply peaked primordial curvature power spectra which, in the presence of non-Gaussian cosmological perturbations, lead to nearly monochromatic PBH mass distributions~\cite{Yoo:2019pma,Matsubara:2022nbr}. Our work however can be easily generalised to account as well for extended PBH mass distributions using the formalism developed in~\cite{Araya:2020tds,Papanikolaou:2023nkx}.
 
We should point out as well here that since we use a simplified effective model in order to capture the MF amplification due to galactic/turbulent dynamo and various instability processes we underestimate the GW amplitude and therefore set conservative constraints on $\Omega_\mathrm{PBH,f}$. Full MHD numerical simulations are required, however, in order account for the convective term in the MF induction equation and account for the aforementioned MF amplifications effects.

Finally, let us highlight that the formalism developed within this work regarding the derivation of the magnetically induced GWs is quite generic and can be applied to any population of magnetised PBHs~\cite{2023arXiv230607804K}, e.g. PBHs with magnetic charge~\cite{Maldacena:2020skw} or Kerr-Newmann PBHs~\cite{Hooper:2022cvr}, thus promoting the portal of MIGWs to a new GW counterpart associated to PBHs, potentially detectable by current/future GW detectors.

\begin{acknowledgments}
T.P. acknowledges financial support from the Foundation for Education and European Culture in Greece as well as the 
contribution of the LISA CosWG and the COST Actions  
CA18108 ``Quantum Gravity Phenomenology in the multi-messenger approach''  and 
CA21136 ``Addressing observational tensions in cosmology with systematics and 
fundamental physics (CosmoVerse)''. This work is also part of the activities of the University of Patras GW group for the LISA consortium.
\end{acknowledgments}

\bibliography{PBH}

\end{document}

%% file: newcommands.tex
\newcommand{\OmegaGW}{\Omega_{\mathrm{GW}}}
\newcommand{\rhoGW}{\rho_{\mathrm{GW}}}

\let\oldsqrt\sqrt
\def\sqrt{\mathpalette\DHLhksqrt}
\def\DHLhksqrt#1#2{%
\setbox0=\hbox{$#1\oldsqrt{#2\,}$}\dimen0=\ht0
\advance\dimen0-0.2\ht0
\setbox2=\hbox{\vrule height\ht0 depth -\dimen0}%
{\box0\lower0.4pt\box2}}

\newcommand{\boldmathsymbol}[1]{{\ensuremath{\boldsymbol{#1}}}}

\newcommand{\calH}{\mathcal{H}}

\newcommand{\beq}{\begin{equation}}
\newcommand{\eeq}{\end{equation}}
\newcommand{\bea}{\begin{equation}\begin{aligned}}
\newcommand{\eea}{\end{aligned}\end{equation}}

\newlength{\wsingfig}
\newlength{\wdblefig}
\newlength{\wquadfig}
\newlength{\wtriplefig}
\newcommand{\Eq}[1]{Eq.~(\ref{#1})}

\newcommand{\Fig}[1]{Fig.~{\ref{#1}}}